\documentclass[%
reprint,
superscriptaddress,
onecolumn,
 amsmath,amssymb,
 aps,
]{revtex4-2}

\usepackage[latin9]{inputenc}
\usepackage{amssymb}
\usepackage{graphicx}
\usepackage{amsfonts}
\usepackage{soul}
\usepackage{color}

\begin{document}
\title{The Gross-Pitaevskii equation for a infinite square-well with a delta-function
barrier}
\author{Robert J. Ragan}
\affiliation{Department of Physics, University of Wisconsin-La Crosse, WI
54601 USA}
\author{Asaad R. Sakhel}
\affiliation{Department of Physics, Faculty of Science, Al-Balqa Applied
University, Salt 19117, Jordan}
\author{William J. Mullin}
\affiliation{Department of Physics, University of Massachusetts, Amherst,
Massachusetts 01003 USA}
\date{\today}

\begin{abstract}
The Gross-Pitaevskii equation is solved by analytic methods for an
external double-well potential that is an infinite square well plus
a $\delta$-function central barrier. We find solutions that have
the symmetry of the non-interacting Hamiltonian as well as asymmetric
solutions that bifurcate from the symmetric solutions for attractive 
interactions and from the 
antisymmetric solutions for repulsive interactions. We present
a variational approximation to the asymmetric state as well as an
approximate numerical approach. We compare with other approximate methods. 
Stability of the states is considered. 
\end{abstract}

\maketitle

\section{Introduction}

The properties of Bose-Einstein condensates in a double-well potential
\cite{Milb}-\cite{Jackson} have received considerable attention
in recent years. The solution of the Gross-Pitaevskii equation (GPE)
with an external double well shows some interesting nonlinear effects,
in particular bifurcations to symmetry-breaking states in both atomic
\cite{Masiello1}-\cite{Mal} and optical systems \cite{Green}-\cite{Hamb}
and unusual dynamics, e.g., self-trapping \cite{Milb,Masiello2,Rag,Ostro,Coullet}.
Symmetry breaking has been observed experimentally in several types
of systems \cite{Green,KevExp,Hamb,ZiboldExp}. In a simple infinite
square well, the equation has exact analytic solutions in terms of
Jacobi elliptical functions \cite{CCRI,CCRII}. When we add a repulsive
Dirac $\delta$-function at the center of the well, we get a double well
that is also soluble exactly. We call this the box-$\delta$  potential.
We can find exact analytic results for the asymmetric solutions as
well. 

 The time dependent GPE is
\begin{equation}
i\hbar\frac{\partial\Phi}{\partial t}=\left(-\frac{\hbar^{2}}{2\bar{m}}\frac{\partial^{2}}{\partial z^{2}}+V_{ex}(z)+g\left|\Phi\right|^{2}\right)\Phi
\end{equation}
with $\Phi$ normalized to $N$; $\bar{m}$ is the particle mass.
Our infinite well potential goes from position $z=-a$ to $a.$ Let
$x=z/a$ and energy be measured in units of $\hbar^{2}/2\bar{m}a^{2}.$
We assume an exponential time dependence for the wave function, $\Phi(x,t)=e^{-i\mu t}\psi(x),$
so the unitless time-independent GPE for the box-$\delta$ potential can
be written in the form
\begin{equation}
\left[-\frac{d^{2}}{dx^{2}}+V(x)+\eta N\left|\psi(x)\right|^{2}\right]\psi(x)=\mu\psi(x),\label{eq:GPE}
\end{equation}
with $\int dx\left|\psi(x)\right|^{2}=1$, $\eta=2\bar{m}a^{2}g/\hbar^{2}$
and 
\begin{equation}
V(x)=\left\{ \begin{array}{cc}
\infty, & \left|x\right|\ge1\\
\gamma\delta(x), & \left|x\right|<1
\end{array}\right.,\label{eq:BoxDel}
\end{equation}
where $\gamma >0$.
 
A potential somewhat similar to what we consider here was a $\delta$-function
within a harmonic potential \cite{AGK}. Refs. \cite{MalPR} - \cite{Mal}
treat the GPE in the box-$\delta$ potential numerically (using a
narrow Gaussian to simulate the $\delta$-function), with approximations
valid for large $\gamma$ and small $\gamma$, and by a variational
approach. We treat it exactly and will compare with results given
in these references.

Sacchetti and Fukuizumi \cite{Sac1,Sac2} have treated a two-level
approximation for an arbitrary two-well potential systematically proving that the results are exact up
to an exponentially small correction in the semiclassical limit ($\hbar\rightarrow\infty)$. We
test this approach as well.

\section{Repulsive interactions}

Here we treat the states that preserve the symmetry of the non-interacting
Hamiltonian; asymmetric states are treated later. Some details of
the Jacobi elliptic functions are given in the Appendix. 

\subsection{Antisymmetric states\label{subsec:Antisymmetric-states}}

\begin{figure}[h]
\centering \includegraphics[width=3in]{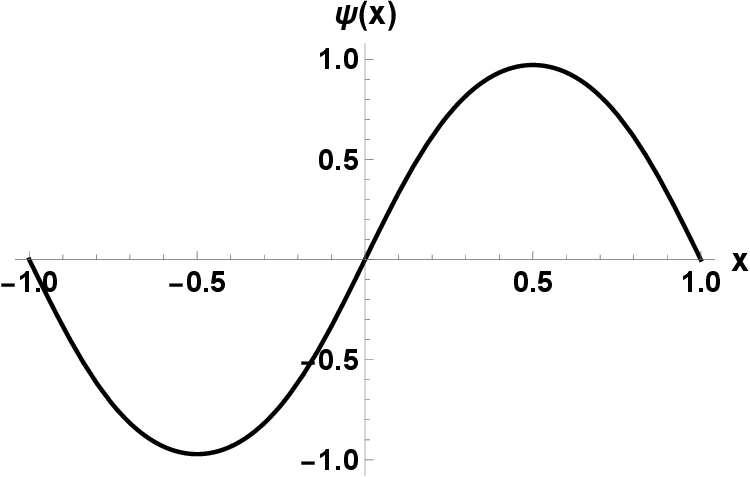}

\caption{The lowest antisymmetric state of the box-$\delta$ well GPE with $\eta N=10$. The state is not affected by the $\delta$-function in the potential.}

\label{figAntisyRep}
\end{figure}

These states vanish at $x=0$ and so are equivalent
to the antisymmetric states of the pure square well \cite{CCRI} with no $\delta$ barrier. 
Substitute the function
\begin{equation}
\psi(x)=A\mathrm{sn}\left(kx|m\right),
\end{equation}
where $\mathrm{sn}$ is the elliptic sine, into Eq. (\ref{eq:GPE})
and equate the coefficients of different powers of $\mathrm{sn}$
to zero to give
\begin{eqnarray}
A^{2} & = & \frac{2k^{2}m}{\eta N},\label{eq:Asq}\\
\mu & = & k^{2}\left(1+m\right).\label{eq:mu}
\end{eqnarray}
Here we see that we are dealing with a repulsive interaction $\eta>0$
because $A^{2}$ is positive. The solution has $\mathrm{sn}\left(0|m\right)=0$,
as it must be for the antisymmetric functions. The boundary conditions
at $x=\pm1$ are satisfied with 
\begin{equation}
k=2jK(m),
\end{equation}
with $j=1,2,\ldots$ and $K(m)$ is the elliptical integral of the
first kind [see Appendix Eq. (\ref{eq:K})]. The normalization integral
is
\begin{equation}
A^{2}\int_{-1}^{1}dx\left(\mathrm{sn}(2jK(m)x|m\right)^{2}=\frac{2A^{2}}{K(m)m}\left[-E(m)+K(m)\right]=1,\label{eq:norm}
\end{equation}
where $E(m)$ is the complete elliptic integral of the second kind
[see Appendix Eq. (\ref{eq:secKnd})]. (This normalization integral
is independent of integer $j.$) The combination of Eqs. (\ref{eq:Asq})
and (\ref{eq:norm}) determine the variable $m,$ which in turn gives
the eigenvalue $\mu$ by Eq. (\ref{eq:mu}). We must solve
\begin{equation}
\frac{16j^{2}K(m)}{\eta N}\left[-E(m)+K(m)\right]=1
\end{equation}
to find $m$. For example, for $j=1$ and $\eta N=10$ we find $m=0.379$
and $\mu=17.16$; for $j=2$, $m=0.1172$, $\mu=46.92$ is the result. Fig. \ref{figAntisyRep} shows an example of this state.

\subsection{Symmetric states}

The symmetric wave function that satisfies the boundary conditions
is
\begin{equation}
\psi(x)=\left\{ \begin{array}{cc}
A\mathrm{sn(}k(x+1)|m), & -1<x<0\\
-A\mathrm{sn}(k(x-1)|m), & 0\le x<1
\end{array}\right..
\end{equation}
This solution satisfies the GPE with the same conditions of Eqs. (\ref{eq:Asq})
and (\ref{eq:mu}). The condition on the wave function derivatives
at $x=0$ is 
\begin{equation}
\Delta\psi^{\prime}(0)=\lim_{\varepsilon\rightarrow0}(\psi^{\prime}(\varepsilon)-\psi^{\prime}(-\varepsilon))=\gamma\psi(0),
\end{equation}
which becomes here
\begin{equation}
2k\mathrm{cn(k|m)\mathrm{dn(k|m)+\gamma\mathrm{sn}(k|m)=0}.}\label{eq:Kcond}
\end{equation}
The normalization becomes 
\begin{equation}
A^{2}\frac{2(k-\mathcal{E}(k|m))}{km}=1,
\end{equation}
where $\mathcal{E}(k|m)$ is the Jacobi Epsilon function (see Appendix
Eq. (\ref{eq:Epsi})). Given the condition of Eq. (\ref{eq:Asq})
we must solve
\begin{equation}
\frac{4k}{\eta N}(k-\mathcal{E}(k|m))-1=0
\end{equation}
simultaneously with that of Eq. (\ref{eq:Kcond}) for $k$ and $m.$

\begin{figure}[h]
\includegraphics[width=6in]{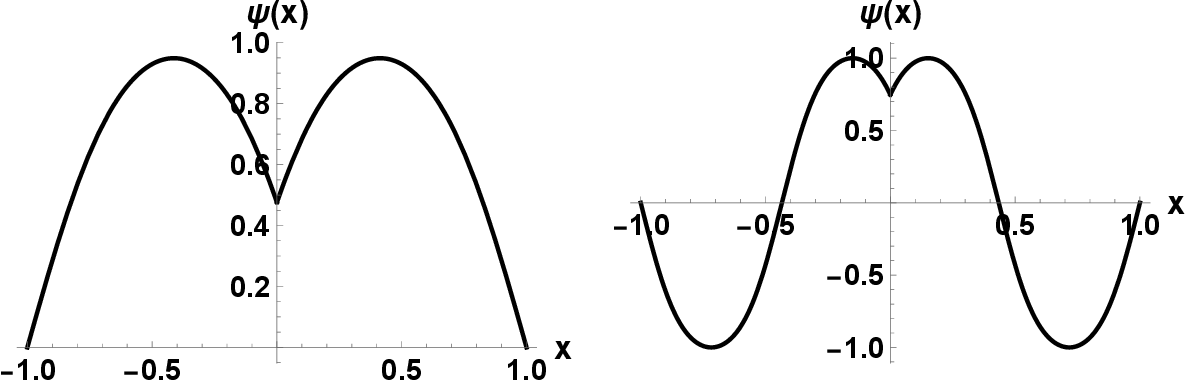}

\caption{The lowest two symmetric states of the box-$\delta$ well GPE with $\gamma=10$
and $\eta N=10.$}

\label{figGPEBoxDelRep}
\end{figure}

For $\gamma=10$ and $\eta N=10,$ we find the solutions for the lowest
energies as 
\begin{eqnarray}
k & = & 3.067;\quad m=0.433;\quad\mu=13.48,\\
k & = & 5.680;\quad m=0.140;\quad\mu=36.77.
\end{eqnarray}
Plots of these are shown in Fig. \ref{figGPEBoxDelRep}. 

\section{Attractive interactions}

Here we again consider the states that maintain the symmetry of the
non-interacting Hamiltonian. 

\subsection{Antisymmetric states\label{subsec:Antisymmetric-states-1}}

As in Sec. \ref{subsec:Antisymmetric-states}, the states are the
same as those where there is no $\delta$-function, which were treated
in Ref. \cite{CCRII}. The wave function
\begin{equation}
\psi(x)=A\mathrm{cn}(kx+d|m),
\end{equation}
where $\mathrm{cn}(x|m)$ is the elliptic cosine, satisfies the GPE
with
\begin{equation}
A^{2}=-\frac{2mk^{2}}{\eta N},\label{eq:AttNorm}
\end{equation}
with $\eta N$ negative for attraction. The other condition is
\begin{equation}
\mu=k^{2}(1-2m).\label{eq:mu2}
\end{equation}
The antisymmetric states have
\begin{eqnarray}
d & = & -K(m),\\
k & = & 2jK(m),
\end{eqnarray}
where $j=1,2,\ldots$ . The normalization equation combined with Eq.
(\ref{eq:AttNorm}) gives the condition
\begin{equation}
2\left[\frac{2\left(2jK(m)\right)^{2}}{\left|\eta N\right|}\right]\frac{E[m]+(m-1)K[m]}{K[m]}=1
\end{equation}
to determine $m$.

\subsection{Symmetric states}

The symmetric states are given by
\begin{equation}
\psi(x)=\left\{ \begin{array}{cc}
A\mathrm{cn(}k(x+1)-K(m)|m), & -1<x<0\\
-A\mathrm{cn}(k(x-1)-K(m)|m), & 0\le x<1
\end{array}\right.
\end{equation}
with again the conditions of Eqs. (\ref{eq:AttNorm}) and (\ref{eq:mu2})
applying to $A$ and $\mu$ in order to satisfy the GPE. The derivative
condition at $x=0$ gives
\begin{equation}
2k\mathrm{sn(}k+K(m)|m)\mathrm{dn(}k+K(m)|m)=\gamma\mathrm{cn(}k+K(m)|m).\label{eq:km1}
\end{equation}
The normalization condition is
\begin{equation}
\frac{2A^{2}}{km\mathrm{dn}(k|m)}\left[k(m-1)\mathrm{dn}(k|m)+\mathrm{dn}(k|m)\mathcal{E}(k|m)-m\mathrm{cn}(k|m)\mathrm{sn}(k|m)\right]=1.
\end{equation}
Thus we must solve Eq. (\ref{eq:km1}) and the following simultaneously
to find $k$ and $m$:
\begin{equation}
\frac{4k}{\left|\eta N\right|\mathrm{dn}(k|m)}\left[k(m-1)\mathrm{dn}(k|m)+\mathrm{dn}(k|m)\mathcal{E}(k|m)-m\mathrm{cn}(k|m)\mathrm{sn}(k|m)\right]=1.
\end{equation}
For $\gamma=10, \eta N=-10,$ we find the lowest two state roots at $\{k,m\}=\{3.085,0.490\}$
and $\{5.651,0.146\}$ corresponding to chemical potentials of 0.187
and 22.602, respectively. 

A wave function here looks similar to that
in the repulsive analysis, except that the amplitude is larger at the peaks and smaller at the $\delta$-function. 

\section{Asymmetric solutions}

The GPE is nonlinear and so has the possibility of developing asymmetric
solutions. The Hamiltonian includes the wave function and so becomes
asymmetric itself when the wave function does. Thus the usual proof
of symmetry or antisymmetry breaks down. 

\subsection{Attractive interactions\label{subsec:Attractive-interactions}}

\begin{figure}[h]
\includegraphics[width=6in]{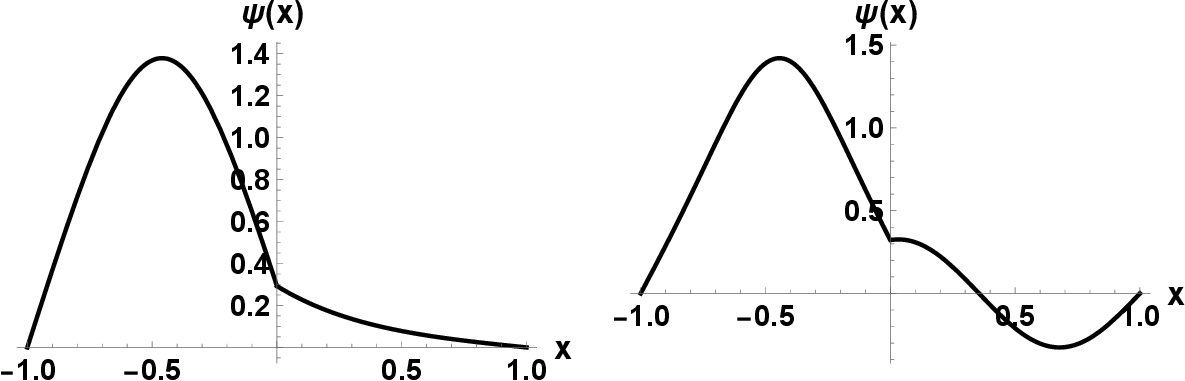}

\caption{Exact analytic asymmetric states of the box-$\delta$ potential GPE
for attractive interactions. (Left) A zero node state 
for $\gamma=10$ and $\eta N=-2.73.$ (Right) A one-node state corresponding
to $\gamma=10$ and $\eta N=-10.2.$ Solutions that are mirror images
of these are degenerate.}

\label{figGPEBoxAsymm}
\end{figure}

For an attractive interaction, the general wave function that satisfies
the boundary conditions at $x=\pm1$, but is not necessarily symmetric
or antisymmetric is 
\begin{equation}
\psi(x)=\left\{ \begin{array}{cc}
A_{1}\mathrm{cn(}k_{1}(x+1)-K(m_{1})|m_{1}), & -1<x<0\\
A_{2}\mathrm{cn}(k_{2}(x-1)-K(m_{2})|m_{2}), & 0\le x<1
\end{array}\right..
\end{equation}
There are six variables to be determined here. The amplitudes and
chemical potentials, as in Sec. \ref{subsec:Antisymmetric-states-1},
obey
\begin{eqnarray}
A_{i}^{2} & = & -\frac{2m_{i}k_{i}^{2}}{\eta N},\label{eq:Ax2}\\
\mu_{i} & = & k_{i}^{2}(1-2m_{i}).
\end{eqnarray}
The wave function must be continuous at $x=0$ giving
\begin{equation}
A_{1}\mathrm{cn(}k_{1}-K(m_{1})|m_{1})=A_{2}\mathrm{cn}(k_{2}+K(m_{2})|m_{2}).
\end{equation}
The chemical potentials must be uniform over the whole system so that
we have 
\begin{equation}
k_{1}^{2}(1-2m_{1})=k_{2}^{2}(1-2m_{2}).\label{eq:mueqn}
\end{equation}
If we integrate in order to satisfy the normalization then we find
\begin{eqnarray}
\frac{A_{1}^{2}}{k_{1}m_{1}\mathrm{dn}(k_{1}|m_{1})}\left[\left(k_{1}(m_{1}-1)+\mathcal{E}(k_{1}|m_{1})\right)\mathrm{dn}(k_{1}|m_{1})-m_{1}\mathrm{cn}(k_{1}|m_{1})\mathrm{sn}(k_{1}|m_{1})\right]+\nonumber \\
+\frac{A_{2}^{2}}{k_{2}m_{2}\mathrm{dn}(k_{2}|m_{2})}\left[\left(k_{2}(m_{2}-1)+\mathcal{E}(k_{2}|m_{2})\right)\mathrm{dn}(k_{2}|m_{2})-m_{2}\mathrm{cn}(k_{2}|m_{2})\mathrm{sn}(k_{2}|m_{2})\right] & = & 1.\label{eq:NormCond}
\end{eqnarray}
The wave function derivative difference at $x=0$ obeys the condition
\begin{eqnarray}
-A_{2}k_{2}\mathrm{sn}(-k_{2}-K(m_{2})|m_{2})\mathrm{dn}(-k_{2}-K(m_{2})|m_{2})+\nonumber \\
+A_{1}k_{1}\mathrm{sn}(k_{1}-K(m_{1})|m_{1})\mathrm{dn}(k_{1}-K(m_{1})|m_{1}) & = & \gamma A_{1}\mathrm{cn}(k_{1}-K(m_{1})|m_{1}).
\end{eqnarray}
These six conditions must be solved simultaneously for $A_{i}$, $k_{i}$,
and $m_{i}$, which can be done, for example, by use of $\mathit{Mathematica's}$
FindRoot command. A ground state and an excited state are shown in
Fig. \ref{figGPEBoxAsymm}.

Note that the main peak in each of these is in the left side; there
are clearly degenerate mirror-image states with the main peak in the
right well. As we will see below these states bifurcate from the symmetric
states at appropriate interaction parameter values. 

\subsection{Repulsive interactions}

For a repulsive interaction, the general wave function that satisfies
the boundary conditions at $x=\pm1$, but is not necessarily symmetric
or antisymmetric is 
\begin{equation}
\psi(x)=\left\{ \begin{array}{cc}
A_{1}\mathrm{sn(}k_{1}(x+1)|m_{1}), & -1<x<0\\
A_{2}\mathrm{sn}(k_{2}(x-1)|m_{2}), & 0\le x<1.
\end{array}\right.
\end{equation}

\begin{figure}[h]
\centering \includegraphics[width=3in]{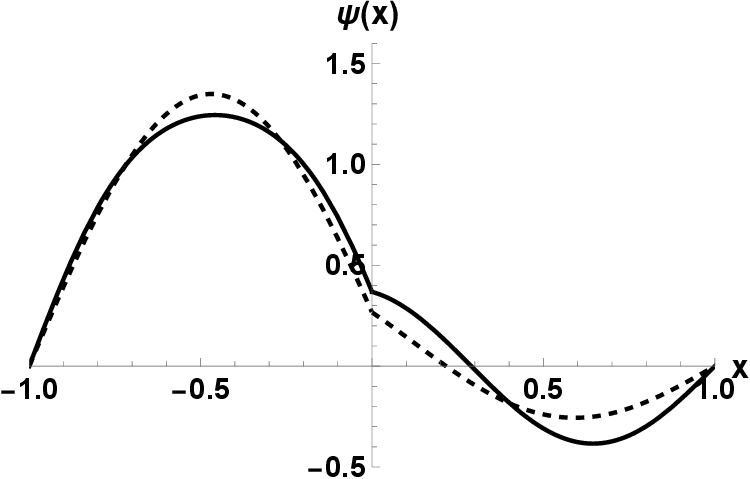}

\caption{(Solid) An asymmetric state solving the GPE for $\gamma=10$ and a
repulsive interaction of $\eta N=10$. (Dashed) The variational state
at the same parameters. Because the state arises from a bifurcation
from antisymmetric states it has one node.}

\label{figRepuls}
\end{figure}

There are again six variables to be determined. The amplitudes and
chemical potentials, as in Sec.\ref{subsec:Antisymmetric-states}
obey
\begin{eqnarray}
A_{i}^{2} & = & \frac{2m_{i}k_{i}^{2}}{\eta N},\label{eq:Amps}\\
\mu_{i} & = & k_{i}^{2}(1+m_{i}).\label{eq:mus}
\end{eqnarray}
We use Eqs. (\ref{eq:Amps}) as the first two conditions. Then Eq.
(\ref{eq:mus}) gives us the condition
\begin{equation}
k_{1}^{2}(1+m_{1})=k_{2}^{2}(1+m_{2}).
\end{equation}
Continuity of the function at $x=0$ implies
\begin{equation}
A_{1}\mathrm{sn(}k_{1}|m_{1})=A_{2}\mathrm{sn}(-k_{2}|m_{2}).
\end{equation}
The derivative condition on the wave function due to the $\delta$-function is
\begin{equation}
A_{2}k_{2}\mathrm{cn}(-k_{2}|m_{2})\mathrm{dn}(-k_{2}|m_{2})-A_{1}k_{1}\mathrm{cn}(k_{1}|m_{1})\mathrm{dn}(k_{1}|m_{1})=\gamma A_{1}\mathrm{sn}(k_{1}|m_{1}).
\end{equation}
The last condition is the normalization, which is
\begin{equation}
\frac{\left[A_{2}^{2}k_{1}k_{2}m_{1}+A_{1}^{2}k_{1}k_{2}m_{2}-A_{1}^{2}k_{2}m_{2}\mathcal{E}(k_{1}|m_{1})-A_{2}^{2}k_{1}m_{1}\mathcal{E}(k_{2}|m_{2})\right]}{k_{1}k_{2}m_{1}m_{2}}=1.
\end{equation}
 These states bifurcate from the repulsive antisymmetric states. As
we try a sequence of $\eta N$ values with $\eta N$ decreasing, the
states that are generated finally evolve from asymmetric into the
antisymmetric states at the bifurcation value. We show a typical repulsive
state in Fig. \ref{figRepuls}. 

\begin{figure}[h]
\centering\includegraphics[width=3in]{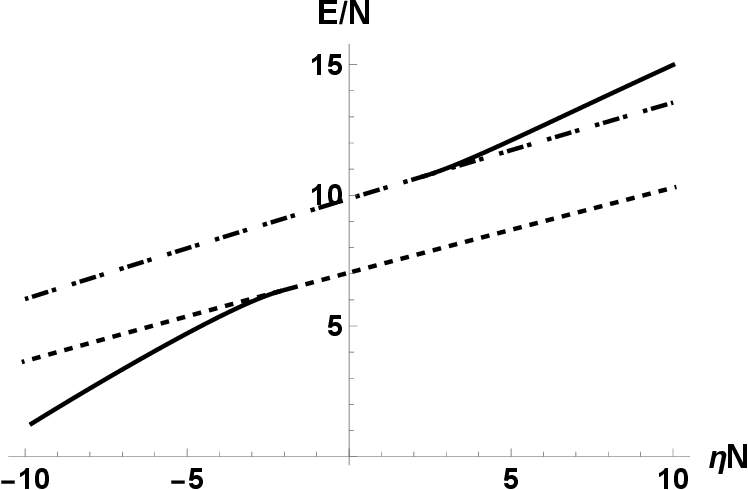}

\caption{Exact energies for the states: (Dashed) symmetric; (dot-dashed) antisymmetric;
(Solid) asymmetric, both attractive and repulsive. The critical attractive bifurcation point of the symmetric state with the asymmetric state is at $\eta_{c}N=-2.07$. For repulsive interaction parameters the bifurcation point for  the asymmetric state energy and the antisymmetric state is $\eta_{c}N=2.34$.  In all cases $\gamma=10.$}

\label{AllEnergy}
\end{figure}

In Fig. \ref{AllEnergy} we summarize the exact results by showing the attractive and repulsive symmetric, antisymmetric and asymmetric energies per particle all on one graph. 

\subsection{Two-mode approximation for asymmetric states \label{subsec:Two-mode-approximation-for}}

We use an approximation method given in Refs. \cite{AGK,Kev2} for
a different potential. The unitless time-dependent GPE is
\begin{equation}
i\frac{\partial\psi}{\partial t}=\left[H_{0}+\eta N\left|\psi\right|^{2}\right]\psi
\label{eq:GPE2}
\end{equation}
where $\psi$ is normalized to $1$ and
\begin{equation}
H_{0}=-\frac{d^{2}}{dx^{2}}+V(x),
\end{equation}
with $V(x)$ being the box-$\delta$ potential. We expand the wave function in a complete set, but truncate
the sum at the first two states: 
\begin{equation}
\psi=\alpha\phi_{0}+\beta\phi_{1}\label{eq:Trial}
\end{equation}
where $\phi_{0}$ is the lowest symmetric eigenfunction of $H_{0}$
with eigenvalue $e_{0},$ and $\phi_{1}$ is the lowest
antisymmetric eigenfunction with eigenvalue $e_{1}$. These
states are
\begin{eqnarray}
\phi_{0}(x) & = & A\left(\sin k\left|x\right|+\frac{2k}{\gamma}\cos kx\right),\label{eq:psi0-1}\\
\phi_{1}(x) & = & \sin\pi x.\label{eq:psi1-1}
\end{eqnarray}
In $\phi_{0}$ the wave number $k$ satisfies 
\begin{equation}
\tan k=-\frac{2k}{\gamma}.
\end{equation}
The energies of these two ideal gas states are, respectively, 
\begin{eqnarray}
e_{0} & = & k^{2},\\
e_{1} & = & \pi^{2}.
\end{eqnarray}
The $\phi_{i}$ are real and are each normalized to unity, and
\begin{equation}
\left|\alpha\right|^{2}+\left|\beta\right|^{2}=1.
\end{equation}
 The equations for the time dependence of the coefficients are 
\begin{eqnarray}
i\dot{\alpha} & = & \alpha e_{0}+\eta N\left[\left|\alpha\right|^{2}\alpha\chi_{40}+\left(\alpha^{*}\beta^{2}+2\left|\beta\right|^{2}\alpha\right)\chi_{22}\right]\\
i\dot{\beta} & = & \beta e_{1}+\eta N\left[\left|\beta\right|^{2}\beta\chi_{04}+\left(\beta^{*}\alpha^{2}+2\left|\alpha\right|^{2}\beta\right)\chi_{22}\right]
\end{eqnarray}
where the dot means first time derivative, and $\chi_{40}=\int dx\phi_{0}^{4}$;
$\chi_{04}=\int dx\phi_{1}^{4}$; $\chi_{22}=\int dx\phi_{0}^{2}\phi_{1}^{2}$.
We look for stationary solutions of the form $\alpha=ue^{-i\Gamma t},$
$\beta=ve^{-i\Gamma t},$ which gives the equations
\begin{eqnarray}
\Gamma u & = & ue_{0}+\eta N\left[u^{3}\chi_{40}+3uv^{2}\chi_{22}\right]\label{eq:Gamma1}\\
\Gamma v & = & ve_{1}+\eta N\left[v^{3}\chi_{04}+3vu^{2}\chi_{22}\right].\label{eq:Gamma2}
\end{eqnarray}
These equations have solutions in the cases where the coefficients
are $u=1$, $v=0$: $\Gamma=e_{0}+\eta N\chi_{40}$, which
is the interacting symmetric state and for $v=1$, $u=0$: $\Gamma=e_{1}+\eta N\chi_{04}$,
the interacting antisymmetric state. $\Gamma$ of Eq. (\ref{eq:Gamma1}) is not the energy, but is a chemical
potential.

To solve more generally multiply (\ref{eq:Gamma1}) by $v$ and (\ref{eq:Gamma2})
by $u$ and subtract the first from the second to give
\begin{equation}
0=\Delta+\eta N\left[v^{2}\chi_{04}-u^{2}\chi_{40}+3\left(u^{2}-v^{2}\right)\chi_{22}\right]
\end{equation}
where
\begin{equation}
\Delta=e_{1}-e_{0}.
\end{equation}
Using $u^{2}+v^{2}=1,$ we can solve for $u^{2}$ giving
\begin{equation}
u^{2}=\frac{\eta N(3\chi_{22}-\chi_{04})-\Delta}{\eta N(6\chi_{22}-\chi_{40}-\chi_{04})}.\label{eq:Extrema}
\end{equation}
We require $0<u^{2}<1$ if we are to have an asymmetric solution.  In the case of $\gamma=10,$ we have $k=2.654$. The energies are
\begin{eqnarray}
e_{0} & = & k^{2}=7.044,\\
e_{1} & = & \pi^{2}=9.870,
\end{eqnarray}
 and we find $\chi_{40}=0.6616;$ $\chi_{04}=$0.75; $\chi_{22}=0.6681$. 

Our condition for an asymmetric solution for the attractive case is
\[
0<\frac{\left|\eta N\right|(3\chi_{22}-\chi_{04})+\Delta}{\left|\eta N\right|(6\chi_{22}-\chi_{40}-\chi_{04})}<1.
\]
All terms here are positive and so the condition on the left is automatic
and the one on the right requires
\begin{equation}
|\eta N|>\frac{\Delta}{3\chi_{22}-\chi_{40}}=2.11.
\end{equation}
We will see in the next section that the exact result is $\eta N_{c}=-2.07$.
The fact that the bifurcation is at $u=1$ implies it is from the
symmetric state. 

In the repulsive case the condition is
\begin{equation}
0<\frac{\eta N(3\chi_{22}-\chi_{04})-\Delta}{\eta N(6\chi_{22}-\chi_{40}-\chi_{04})}<1. \label{eq:crit}
\end{equation}
 The critical limit is the left side, which requires
\begin{equation}
\eta N_{c}=\frac{\Delta}{3\chi_{22}-\chi_{04}}=2.25.
\end{equation}
We will see in the next section that the exact result is 2.34. It
is a bifurcation at $u\sim 0$, that is, from the antisymmetric state. 

\begin{figure}[h]
\centering\includegraphics[width=3in]{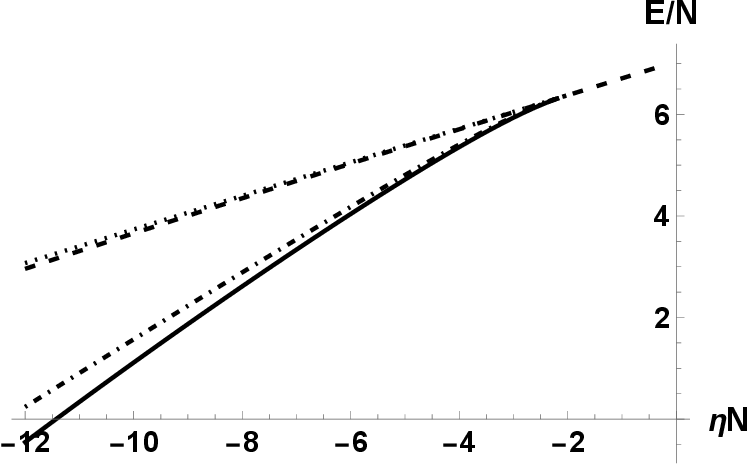}

\caption{Variational results for asymmetric states with attractive interactions
and $\gamma=$10: (Dashed line) The exact energy per particle as a
function of interaction $\eta N$ for the symmetric state, and (solid
line) the exact asymmetric state. (Dotted line--barely above the
exact dashed line) the variational symmetric state; and (dash-dotted
line) that for the asymmetric state. The variational bifurcation
occurs at $\eta N=-2.3$ compared to the exact result at$-2.07.$}

\label{figEnergies}
\end{figure}

 The variational energy for the trial wave function of Eq.
(\ref{eq:Trial}) is
\begin{equation}
E=u^{2}e_{0}+v^{2}e_{1}+\frac{\eta N}{2}\left(\chi_{40}u^{4}+\chi_{04}v^{4}+6\chi_{22}u^{2}v^{2}\right).\label{eq:TrialE}
\end{equation}
The extrema of this energy are given by Eq. (\ref{eq:Extrema}). For
the attractive case this is a minimum, but for the repulsive case
it is a maximum. This is an example of a state with static instability
but dynamic stability \cite{JKL}. A variational state and the exact
wave function are compared in Fig. \ref{figRepuls}. The comparison
between exact energies and variational energies in the attractive case is given in Fig. \ref{figEnergies}.

For very large interaction $\eta N$, where the kinetic energy and
the external potential become numerically unimportant, the parameter
$u$ of Eq. (\ref{eq:Extrema}) approaches a constant--the wave
function no longer changes!

\subsection{Sacchetti two-mode method}\label{subsec:Sacchetti-method}

Refs. \cite{Sac1,Sac2} develop a two-state approximation for any
two-well potential, which they show is exponentially accurate in the
semiclassical limit of $H_{0}$, which in our case corresponds to
$\gamma\rightarrow\infty$. Their two basis states are mostly localized
in the right or left well, that is, in our case,
\begin{eqnarray}
\phi_{R} & = & \frac{1}{\sqrt{2}}\left(\phi_{0}+\phi_{1}\right),\\
\phi_{L} & = & \frac{1}{\sqrt{2}}\left(\phi_{0}-\phi_{1}\right).
\end{eqnarray}
Overlap integrals of the states become small as $\gamma\rightarrow\infty$.
Define 
\begin{eqnarray}
\Omega & = & \frac{1}{2}\left(e_{1}+e_{0}\right),\\
\omega & = & \frac{1}{2}\left(e_{1}-e_{0}\right)=\frac{\Delta}{2}.
\end{eqnarray}
Their expansion of the asymmetrical state is
\begin{equation}
\psi=e^{-i\Omega t}\left(a_{R}\phi_{R}+a_{l}\phi_{L}\right).
\end{equation}
Substituting this in Eq. (\ref{eq:GPE2}) and dropping all overlap
integrals between $\phi_{R}$ and $\phi_{L}$ gives
\begin{eqnarray}
i\dot{a}_{R} & = & -\omega a_{L}+\eta N\left|a_{R}\right|^{2}a_{R}W_{40},\\
i\dot{a}_{L} & = & -\omega a_{R}+\eta N\left|a_{L}\right|^{2}a_{L}W_{04},
\end{eqnarray}
with 
\begin{equation}
W_{nm}=\int dx\phi_{R}^{n}\phi_{L}^{m}=W_{mn}.
\end{equation}
Define the phases of $a_{R}$ and $a_{L}$ as $\alpha$ and $\beta$
with new variables 
\begin{eqnarray}
z & = & \left|a_{R}\right|^{2}-\left|a_{L}\right|^{2},\\
\theta & = & \alpha-\beta.
\end{eqnarray}
The variable $z$ is an asymmetry parameter, being $+1$
if the wave function is all in the right well and $-1$ if it is in
the left well. They find the results
\begin{eqnarray}
\dot{\theta} & = & \frac{-2z}{\sqrt{1-z^{2}}}\cos\theta-\zeta z,\\
\dot{z} & = & 2\sqrt{1-z^{2}}\sin\theta,\label{eq:zdot}
\end{eqnarray}
where the effective particle interaction parameter is
\begin{equation}
\zeta=\frac{W_{40}}{\omega}\eta N.
\end{equation}
 For stationary solutions $\dot{\theta}=0$, $\dot{z}=0$, Eq. (\ref{eq:zdot})
implies $\theta=0$ or $\pi$; with these we can have the solution $z=0$
for each, corresponding respectively to symmetric and antisymmetric solutions.
For asymmetric solutions we want $z\neq0$. For $\theta=0$ or $\pi$
this means 
\begin{equation}
f_{\pm}(z)=\frac{\mp2z}{\sqrt{1-z^{2}}}-\zeta z=0.
\end{equation}
If $\zeta>0$ then the upper sign has only $z=0$ as a solution; the
lower sign gives
\begin{equation}
z=\pm\sqrt{1-\frac{4}{\zeta^{2}}}\label{eq:z}.
\end{equation}
This result has $z=0$ at the bifurcation point 
\begin{equation}
\zeta_{c}=2.
\end{equation}

\begin{figure}[h]
\centering \includegraphics[width=3in]{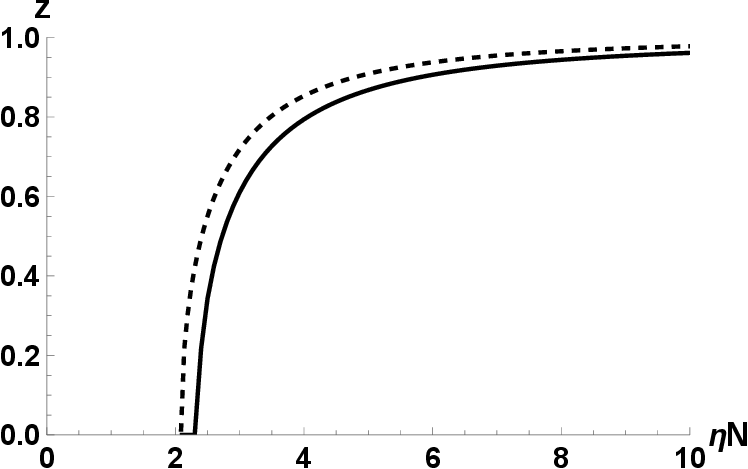}
\caption{The asymmetry parameter, $z_{ex}$ of Eq. (\ref{eq:zex}) (solid) for
$\gamma=10,$ compared to the two-level semiclassical $z$ of Eq. (\ref{eq:z})
(dashed) versus $\eta N$.}
\label{fzplot}
\end{figure}

We can compare this result 
for the asymmetry variable $z$ with our
exact results by calculating the equivalent of $\left|a_{R}\right|$
and $\left|a_{L}\right|$, namely
\begin{eqnarray}
A_{R} & = & \int dx\phi_{R}\psi_{as},\\
A_{L} & = & \int dx\phi_{L}\psi_{as},
\end{eqnarray}
where $\psi_{as}$ is an exact asymmetrical state, as we found it.
What we find for the repulsive asymmetrical states is that $A_{R}^{2}+A_{L}^{2}>0.997$
for all $\eta N$ considered; the two level approximation seems quite
good! We then define the related asymmetry for the exact state as
\begin{equation}
z_{ex}=\frac{A_{L}^{2}-A_{R}^{2}}{A_{L}^{2}+A_{R}^{2}}.\label{eq:zex}
\end{equation}
The included renormalization makes negligible change. A plot
comparing the semiclassical $z$ of Eq. (\ref{eq:z}) and $z_{ex}$
is shown in Fig. \ref{fzplot}.

\subsection{Approximate numerical method\label{subsec:Approximate-numerical-method}}

\begin{figure}[h]
\centering\includegraphics[width=3.5in]{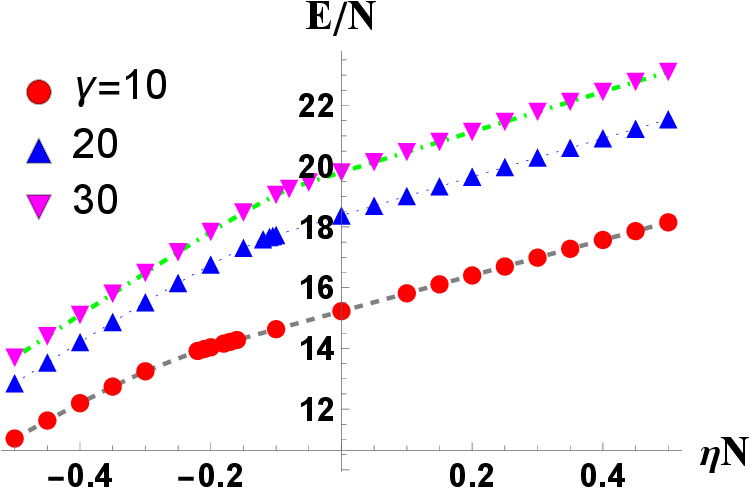}

\caption{MCTDHX energy per particle E/N versus the interaction parameter $\eta N$
for a number of $\delta$-function strengths $\gamma$. The number of
particles is $N=10$. The kink in the graphs gives the bifurcation
critical point. The time step taken is usually adjusted by the code
itself and is 0.0078125 (unitless) for the present case; the simulation
grid has 256 pixels; and the runs are taken over an imaginary time
from $t=0$ to 30 (unitless). The parameters of Eqs. (\ref{eq:Gaus})
and (\ref{eq:wall}) are $\xi=0.05$, $p=1000$, and $L=0.495$.}

\label{figMCTDHX}
\end{figure}

Numerical solutions of the GPE are, of course, also possible. Such
an approach was described for the box-$\delta$ external potential in
Ref. \cite{MalPR}. We have used the so-called multi-configurational
time-dependent Hartree method (MCTDHX) for bosons \cite{MCTDHX}.
This method solves the many-body time-dependent Schr$\mathrm{\ddot{o}}$dinger
equation using variational approaches, and is very accurate. The method
also allows multiple states to contain condensates. By setting the
number of orbitals equal to one (i.e., only the ground state condensate
is allowed) one equivalently solves the GPE.
It is possible to obtain static as well as dynamic solutions. We simulate
the box-$\delta$ potential by a Gaussian, as was done in Ref. \cite{MalPR}:
\begin{equation}
V(x)=\frac{\gamma}{\sqrt{\pi}\xi}e^{-x^{2}/\xi^{2}}\label{eq:Gaus}
\end{equation}
with the infinite box walls simulated by a power-law trap:
\begin{equation}
V_{ex}(x)=\left|\frac{x}{L}\right|^{p}.\label{eq:wall}
\end{equation}
Here $L$ is a length scale and $p$ is an exponent that
generates flatness around $x=0$ for $p\gg1$, $\xi$ is a measure
of the width of the Gaussian, and $\gamma$ is the strength parameter
we used for the $\delta$-function barrier. To find bifurcation points
we solved the time-dependent equations in imaginary time, which gives
the ground state, for a range of parameters $\eta$ and $\gamma$
while keeping the number of particles fixed at $N=10.$ The results are presented in Fig. \ref{figMCTDHX}. The critical
bifurcation point is determined by the position of the kink in the
energy. 

\subsection{Bifurcation\label{subsec:Bifurcation}}

In nonlinear equations fixed points often bifurcate; one solution
may become unstable while a separate solution becomes stable at that
point. Fig. \ref{figEnergies} shows the energy per particle for
the symmetric and the asymmetric states with the latter separating
from the former at $\eta N=-2.07.$ As we step through increasing
$\eta N$ values looking for asymmetric parameters, we find that the
asymmetric state evolves into the symmetric state in the attractive
case precisely at the critical $\eta N$ value. That is, near the
critical value the state is almost symmetric, with nearly identical
double humps, i.e., asymmetry parameter $z=0$ there. Note that in the attractive
case the energy for the asymmetric state is lower than the corresponding
symmetric state, but in the repulsive case it is higher than that
of the antisymmetric state. However, that statement does not address
the dynamical stability of the states. 

We can compare several methods of computing the bifurcations points:

\subsubsection{Variational estimate of bifurcation points}

In Sec. \ref{subsec:Two-mode-approximation-for} we showed that the
variational method gave the critical $\eta$ value for attractive
interactions as
\begin{equation}
\left(\eta N\right)_{c}^{S}=\frac{-\Delta}{3\chi_{22}-\chi_{40}}.\label{eq:Vari}
\end{equation}
The superscript $S$ notes that the bifurcation is with the symmetric
state. For repulsive interactions, we have
\begin{equation}
\left(\eta N\right)_{c}^{AS}=\frac{\Delta}{3\chi_{22}-\chi_{04}}\label{eq:Antis}.
\end{equation}
The bifurcation is with the antisymmetric state. These results are
plotted in Fig. \ref{malomed} as a function of $\gamma$.

\subsubsection{Numerical energy computation}

The method described in Sec. \ref{subsec:Approximate-numerical-method}
finds critical bifurcation points in good agreement with the exact
values and these are also shown in Fig. \ref{malomed}.

\subsubsection{Estimates of Refs. \cite{MalPR} and \cite{Mal}}

\begin{figure}[h]
\includegraphics[width=3in]{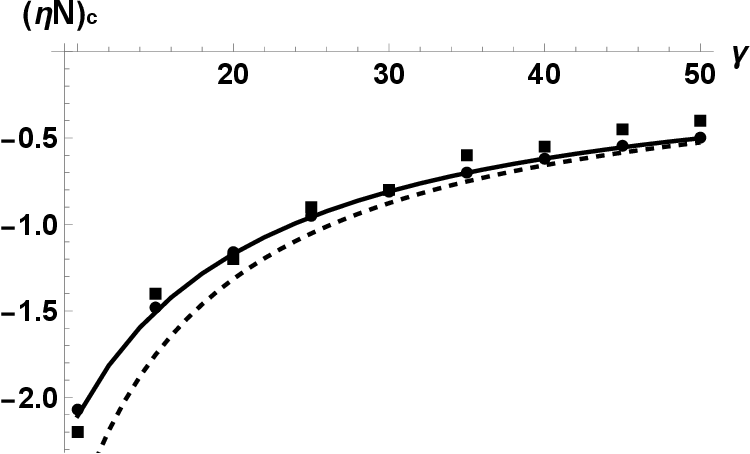}$\quad\quad$\includegraphics[width=3in]{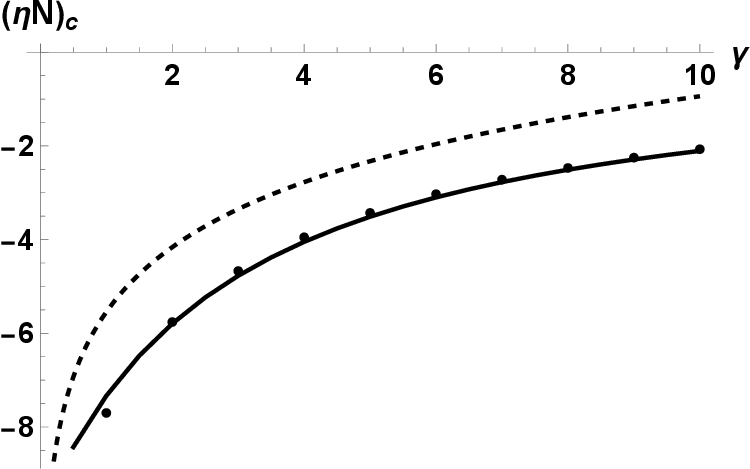}

\caption{(Left) Values of the critical interaction parameter $\left(\eta N\right)_{c}$ for
large $\gamma$ with attractive interactions: (dots) exact computations;
(squares) MCTDHX numerical method; (solid line) Eq. (\ref{eq:Vari});
(dashed line) Eq. (\ref{eq:Malo}). (Right) Values of the critical
parameter for small $\gamma$ and attractive interactions: (dots)
exact computations; (solid line) Eq. (\ref{eq:Vari});(dashed line)
Eq. (\ref{eq:MalSmall}).}

\label{malomed}
\end{figure}

Refs. \cite{MalPR} and \cite{Mal} consider finding a value of the
critical bifurcation interaction parameter for large $\gamma$ by
considering approximations for the wave function and for the interaction
term. They predict that the critical interaction parameter will
be given by
\begin{equation}
\left(\eta N\right)_{c}^{C}=\frac{8\pi^{2}}{3\gamma}.\label{eq:Malo}
\end{equation}
This estimate is said to be valid for $\gamma\gg6.$ These authors give
a treatment valid for small $\gamma$ resulting in the formula 
\begin{equation}
\left(\eta N\right)_{c}^{D}=2\ln\left(16/\gamma\right).\label{eq:MalSmall}
\end{equation}
Ref. \cite{MalPR} also gives a numerical treatment with an Gaussian
approximation for the barrier and a general variational treatment.
In Fig. \ref{malomed} (Left) we compare our exact values with the MCTDHX numerical
results (squares) and the estimates of Eqs. (\ref{eq:Vari})
and (\ref{eq:Malo}). Fig. \ref{malomed} (Right) shows exact results
compared with Eqs. (\ref{eq:Vari}) and (\ref{eq:MalSmall}).

\subsubsection{Sacchetti estimates}

\begin{figure}[h]
\centering\includegraphics[width=3in]{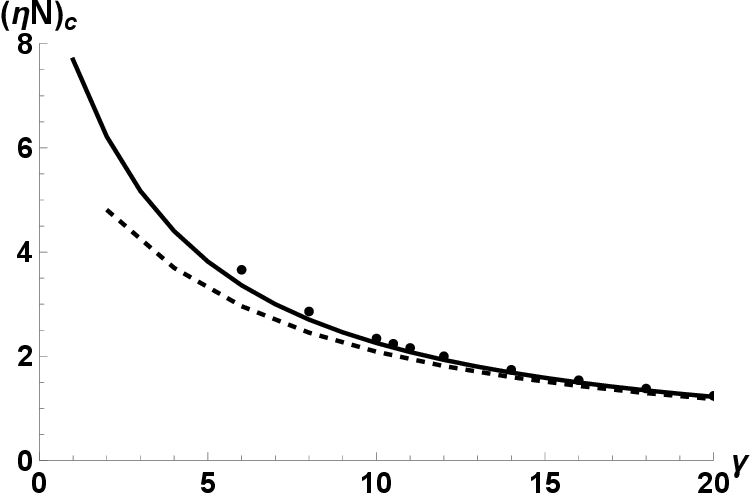}

\caption{Values of the repulsive critical interaction parameter $\left(\eta N\right)_{c}$ versus
$\gamma$ for the repulsive antisymmetrical state: (dots) exact computations;
(solid line) Eq. (\ref{eq:Antis}); (dashed) Eq. (\ref{eq:Sacetac}).}

\label{figCritRepul2}
\end{figure}

In Sec. \ref{subsec:Sacchetti-method} we found the asymmetry parameter
$\xi_{c}=2$ which translates to 
\begin{equation}
\eta N_{c}=\frac{2\omega}{W_{40}}.\label{eq:Sacetac}
\end{equation}
Fig. \ref{figCritRepul2} gives the equivalent results for repulsive
interactions where the antisymmetric state bifurcates to an asymmetric
wave function. It shows the exact bifurcation points, those of Eq.
(\ref{eq:Antis}) from the variational estimate, and the semiclassical
result of Sacchetti, Eq. (\ref{eq:Sacetac}). One sees that the
semiclassical result is quite accurate by $\gamma=20.$

\section{Dynamical Stability\label{subsec:Relative-Stability}}

In order to determine the stability of the GPE stationary solutions
we carry out a Bogoliubov-type stability analysis as described in
Ref. \cite{MalPR}. We consider perturbations of the form 
\begin{equation}
\psi(x,t)=\left[\psi_{i}(x)+\delta\psi(x,t)\right]e^{-i\mu t}\label{eq:Pertur}
\end{equation}
where $\psi_{i}(x)$ is the unperturbed wave function at $t=0$, $\delta\psi=u(x)e^{-i\lambda t}+v^{*}(x)e^{i\lambda^{*}t}$;
$u$ and $v$ are eigenmodes, and $\lambda$ is the corresponding
eigenfrequency. The stability criterion is then $\mathrm{Im}[\lambda]=0$.  The method we use to 
solve the GPE with this wave function is described in Appendix B.
In Ref. \cite{MalPR} the stability of the symmetric mode was investigated
for attractive interactions. Our results for $\gamma=10$ are similar;
the symmetric solution becomes unstable beyond the bifurcation point $\eta N\leq-2.07$.
Stability analysis of the asymmetric
state shows it to be stable, as it should be a standard pitchfork bifurcation. We
find no instabilities of the symmetric groundstate with repulsive interactions.

\begin{figure}[h]
\centering \includegraphics[width=3in]{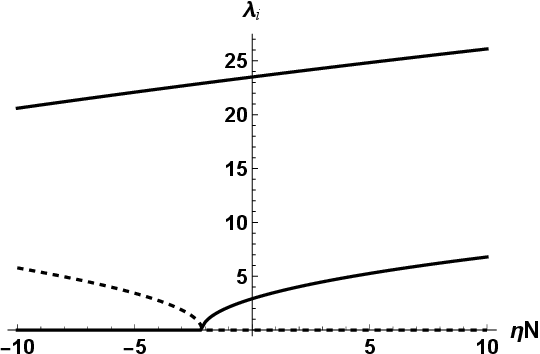}

\caption{Real (solid) and imaginary (dashed) parts of the two lowest eigenfrequencies
of Eq. (\ref{eq:Pertur}) for the symmetric GPE state with $\gamma=10$.
The symmetric state is unstable for $\eta N\le-2.07$. Only the
positive parts are shown, the negative parts are identical.}

\label{Rag1}
\end{figure}

\begin{figure}[h]
\centering\includegraphics[width=3in]{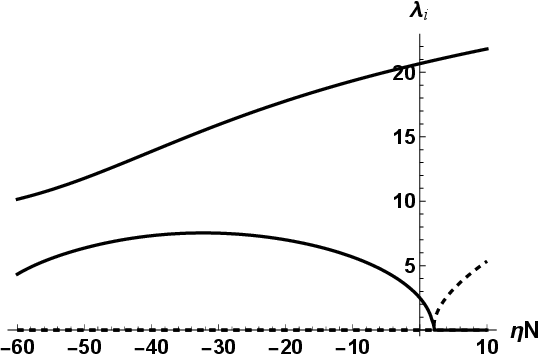}

\caption{Same as Fig. \ref{Rag1} for the antisymmetric state, also with $\gamma=10$.
The antisymmetric state becomes unstable for $\eta N>2.34$. }

\label{Rag2}
\end{figure}

\begin{figure}[h]
\centering\includegraphics[width=3in]{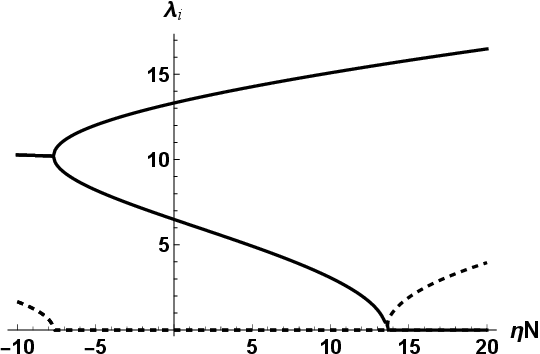}

\caption{Eigenfrequencies of the antisymmetric state for $\gamma=1$. At $\eta N=-7.6$
the first and second eigenfrequencies coalesce into complex conjugate
pairs, corresponding to oscillatory instability with real frequencies
$\pm10.4$. }

\label{Rag3}
\end{figure}

Similarly, we find for $\gamma=10$ the antisymmetric state is unstable for $\eta N$ 
greater than the bifurcation threshold $+2.34$.  For $\gamma = 10$ we find no oscillatory 
instability for attractive interactions. This would seem to differ from Ref. \cite{MalPR}, 
except their reported results were for $\gamma\leq4$.  When we repeat the analysis for 
$\gamma=1$ we find oscillatory instability at $\eta N<-7.6$  in agreement with their results, 
and non-oscillatory instability at $\eta N>13.5$. See Figs. \ref{Rag1}-\ref{Rag3}. In Fig. \ref{Rag3}
the eigenvalues coalesce in contrast to those in Fig. \ref{Rag2}. There is a value of $\gamma$ for 
which the two eigenvalues just touch ($\gamma=5.94$).
Refs. \cite{Sac1,Sac2} do not report oscillatory instabilities because this only 
happens for small gamma, i.e. large tunneling, where the semiclassical two-mode 
approximation is not valid.  Furthermore the oscillatory instability requires at 
least 3 modes, as explained in Appendix B.

\begin{figure}[h]
\includegraphics[width=3.3in,bb = 68 468 526 703,clip]{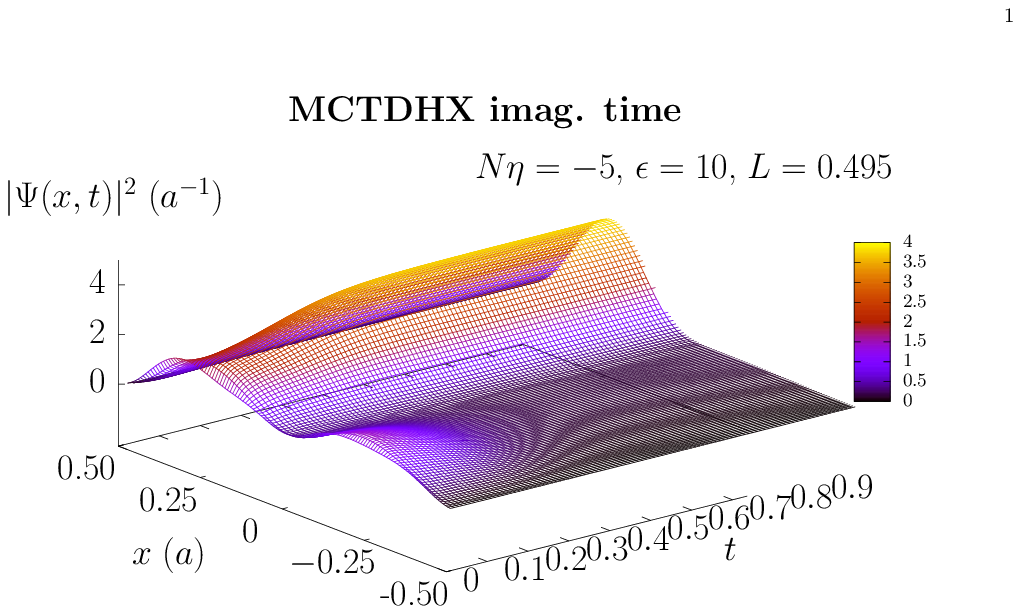}
$\quad\quad$\includegraphics[width=2.7in,bb = 0 1 444 313,clip]
{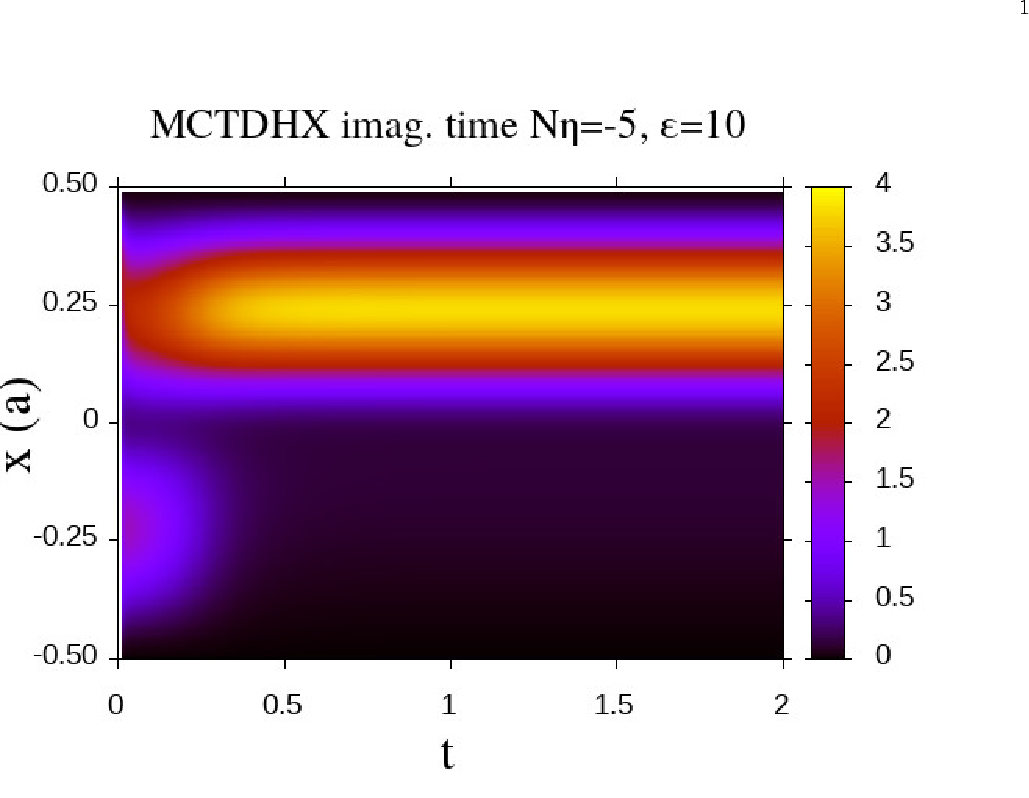}

\caption{(in color online)(Left) Evolution in time from an unstable symmetric wave function
to a stable asymmetric function for $\eta N=-5$, $\gamma=10$ as computed
by the MCTDHX imaginary time method; (Right) an alternative view of the same evolution. 
The MCTDHX time step, simulation grid, $\xi$, $p$ and $L$ are the same as in Fig. \ref{figMCTDHX}. }

\label{Evol}
\end{figure}

Another way of looking at stability is to solve the time dependent
GPE with an unstable initial wave function and watch the result evolve
to a stable state. We show that in Figs. \ref{Evol} (Left)  where
the unstable symmetric state evolves into a stable asymmetric state.
This is repeated in a different form in Fig. \ref{Evol} (Right).

\section{Discussion}

We solved the Gross-Pitaevskii equation for the simple box-$\delta$ external potential analytically for attractive and repulsive interactions in terms of Jacobi elliptic functions. We find
symmetric, antisymmetric, and asymmetric functions become the stable states for interaction parameters in appropriate parameter regions. We were able to demonstrate bifurcation from a symmetric state to the asymmetric state below a critical attractive interaction parameter, and from an antisymmetric to an asymmetric state above a critical repulsive interaction. A two-mode variational approximation gives accurate estimates of the critical parameters.  An approximate numerical method, the MCTDHX, has also been used to make comparisons between numerical and analytical results and agrees well with the exact results. Comparison demonstrates previous approximate treatments are quite accurate. A stability analysis shows that the symmetric state becomes unstable below the attractive bifurcation critical point and the antisymmetric state above the repulsive critical point. The antisymmetric state shows an oscillatory instability for attractive interactions within a range of the parameter space. The asymmetric states are stable. We find the theorems of Refs. \cite{Sac1,Sac2} relating to bifurcation and stability of states in a general two-well system are accurate in our exact case. 

The system can be generalized, for example, to non-symmetrical external potentials 
for which the bifurcation type may change \cite{Kev2}.  In a subsequent paper \cite{SRM} we will show the box-$\delta$ potential to be quite convenient for the study of the
accuracy of the GPE in comparison with the more general second-quantized
Fock Schr$\mathrm{\ddot{o}}$dinger equation. 

\section{Acknowledgements}

We thank Profs. Boris Malomed and Panos Kevrekidis for very useful
comments. We also thank Paolo Molignini, Rui Lin, and Axel Lode for use of the MCTDHX code. The MCTDHX calculations were performed at the PARADOX supercomputing facility of the Scientific Computing Laboratory (SCL), Institute of Physics Belgrade, Serbia.

%\clearpage{}

\section{Appendix A: Jacobi elliptic functions}

We will need the three Jacobi functions $\mathrm{sn}(x|m)$, $\mathrm{cn}(x|m),$
and $\mathrm{dn}(x|m)$ where $y=\mathrm{dn}(x)$ satisfies
\begin{equation}
\frac{d^{2}y}{dx^{2}}+2y^{3}-(2-m)y=0;
\end{equation}
 $y=\mathrm{sn}(x|m)$ satisfies
\begin{equation}
\frac{d^{2}y}{dx^{2}}-2my^{3}+(1+m)y=0;\label{eq:sneq}
\end{equation}
and $y=\mathrm{cn}(x|m)$ satisfies 
\begin{equation}
\frac{d^{2}y}{dx^{2}}+2my^{3}+(1-2m)y=0.\label{eq:cneq}
\end{equation}
We also have limits like
\begin{equation}
\mathrm{sn}(x|0)=\sin(x),\mathrm{\qquad sn}(x|1)=\tanh(x),
\end{equation}
 and the relations
\begin{eqnarray}
\frac{d\mathrm{sn}(x|m)}{dx} & = & \mathrm{cn}(x|m)\mathrm{dn}(x|m),\\
\frac{d\mathrm{cn}(x|m)}{dx} & = & -\mathrm{sn}(x|m)\mathrm{dn}(x|m),\\
\frac{d\mathrm{dn}(x|m)}{dx} & = & -m\mathrm{sn}(x|m)\mathrm{cn}(x|m),\\
\mathrm{sn}(x|m)^{2}+\mathrm{cn}(x|m)^{2} & = & 1,\\
\mathrm{dn}(x|m)^{2} & = & 1-m\mathrm{sn}(x|m)^{2}.
\end{eqnarray}
Another important quantity is $K(m),$ the complete elliptic integral
of the first kind:
\begin{equation}
K(m)=\int_{0}^{\pi/2}\frac{d\phi}{\sqrt{1-m\sin^{2}\phi}},\label{eq:K}
\end{equation}
where $0<m<1$, which determines the period of the Jacobi functions as
in 
\begin{equation}
\mathrm{sn}(x+2nK(m)|m)=(-1)^{n}\mathrm{sn}(x|m),\label{eq:per}
\end{equation}
where $n$ is an integer. The elliptic integral of the second kind
is also used above:
\begin{equation}
E(m)=\int_{0}^{\pi/2}d\phi\sqrt{1-m\sin^{2}\phi}.\label{eq:secKnd}
\end{equation}
The Jacobi epsilon function has the integral representation
\begin{equation}
\mathcal{E}(m)=\int_{0}^{\mathrm{sn}(x|m)}\sqrt{\frac{1-m^{2}t^{2}}{1-t^{2}}}.\label{eq:Epsi}
\end{equation}

\section{Appendix B: Stability calculation details}

We wish to test the stability of the solution $\Phi_{0}(x,t)$ to
the time dependent GPE
\begin{equation}
i\hbar\frac{\partial\Phi_{0}}{\partial t}=\left(H_{0}+\eta N\left|\Phi_{0}\right|^{2}\right)\Phi_{0},
\end{equation}
 with the noninteracting Hamiltonian
\begin{equation}
H_{0}=-\frac{d^{2}}{dx^{2}}+V(x).
\end{equation}
We substitute the form
\begin{equation}
\Phi_{0}(x,t)\rightarrow\left[\Phi_{0}(x)+u(x)e^{-i\lambda t}+v^{*}(x)e^{i\lambda^{*}t}\right]e^{-i\mu t},\label{eq:PSI}
\end{equation}
where $\mu$ is the chemical potential and $u$ and $v$ small perturbations to the solution $\Phi_{0}(x)$.
The stability criterion is that $\mathrm{Im}[\lambda]=0.$ We maintain
only first order in the perturbations $u$ and $v$ and use the zero
order equation
\begin{equation}
\left(H_{0}+\eta N\Phi_{0}^{2}\right)\Phi_{0}=\mu\Phi_{0}.
\end{equation}
The quantities $e^{-i\lambda t}$ and $e^{i\lambda^{*}t}$ are linearly
independent and that gives the Bogoliubov equations
\begin{eqnarray}
H_{0}u+\eta N\Phi_{0}^{2}(2u+v) & = & \left(\mu+\lambda\right)u,\\
H_{0}v+\eta N\Phi_{0}^{2}(2v+u) & = & \left(\mu-\lambda\right)v,
\end{eqnarray}
Adding these two equations with $s=u+v$ and $t=u-v$ gives 
\begin{eqnarray}
\left(H_{0}-\mu\right)s+3\eta N\Phi_{0}^{2}s & = & \lambda t\label{eq:Hs},\\
\left(H_{0}-\mu\right)t+\eta N\Phi_{0}^{2}t & = & \lambda s\label{eq:Ht}.
\end{eqnarray}
These equation can be solved numerically (e.g., \cite{MalPR}) by
substituting a Gaussian for the $\delta$-function barrier, but here
we present an alternative approach that does not use that approximation.

We will solve these equations by expanding $u$ and $v$ in the complete
set of states of the noninteracting Hamiltonian:
\begin{equation}
H_{0}\phi_{\alpha}=\varepsilon_{\alpha}\phi_{\alpha}.
\end{equation}
We have
\begin{eqnarray}
u(x) & = & \sum_{\alpha}u_{\alpha}\phi_{\alpha},\label{eq:Uexp}\\
v(x) & = & \sum_{\alpha}v_{\alpha}\phi_{\alpha}.\label{eq:Vexp}
\end{eqnarray}
We use the same symbol, as with $u$, to represent the function $u(x)$
and the expansion coefficient $u_{\alpha}$ but these are always distinguished
by the subscript on the coefficient. The expansion gives
\begin{eqnarray}
u_{\alpha}\varepsilon_{\alpha}+\sum_{\beta}\Omega_{\alpha\beta}\left(2u_{\beta}+v_{\beta}\right) & = & \left(\mu+\lambda\right)u_{\alpha},\\
v_{\alpha}\varepsilon_{\alpha}+\sum_{\beta}\Omega_{\alpha\beta}\left(u_{\beta}+2v_{\beta}\right) & = & \left(\mu-\lambda\right)v_{\alpha},
\end{eqnarray}
where
\begin{equation}
\Omega_{\alpha\beta}=\eta N\left\langle \alpha\right|\left|\Phi_{0}\right|^{2}\left|\beta\right\rangle .
\end{equation}
Take the sum and difference of the two equations with 
\begin{eqnarray}
s_{\alpha} & = & u_{\alpha}+v_{\alpha}\label{eq:s},\\
t_{\alpha} & = & u_{\alpha}-v_{\alpha}\label{eq:t},
\end{eqnarray}
to find
\begin{eqnarray}
M_{3}s & = & \lambda t\label{eq:M3s},\\
M_{1}t & = & \lambda s\label{eq:M1t},
\end{eqnarray}
where 
\begin{eqnarray}
M_{3\alpha\beta} & = & \left(\varepsilon_{\alpha}-\mu\right)\delta_{\alpha\beta}+3\Omega_{\alpha\beta},\\
M_{1\alpha\beta} & = & \left(\varepsilon_{\alpha}-\mu\right)\delta_{\alpha\beta}+\Omega_{\alpha\beta}.
\end{eqnarray}
We then can find the eigenvalues by using the product matrix:
\begin{equation}
M_{1}M_{3}s=\lambda^{2}s\label{eq:EigEq},
\end{equation}
or alternatively
\begin{equation}
M_{3}M_{1}t=\lambda^{2}t.\label{eq:tEq}
\end{equation}
 $M_{1}$ and $M_{3}$ are each Hermitian, but $M_{3}M_{1}$and $M_{1}M_{3}$
are not.

We have, from Eqs. (\ref{eq:s}) and (\ref{eq:t}),
\begin{eqnarray}
u_{\alpha} & = & \frac{1}{2}\left(s_{\alpha}+t_{\alpha}\right),\\
v_{\alpha} & = & \frac{1}{2}\left(s_{\alpha}-t_{\alpha}\right).
\end{eqnarray}
From these coefficients we can find the spatial states $u(x)$ and
$v(x)$ by Eqs. (\ref{eq:Uexp}) and (\ref{eq:Vexp}). We have
\begin{equation}
\sum_{\alpha}\left(u_{\alpha}^{2}-v_{\alpha}^{2}\right)=\sum_{\alpha}s_{\alpha}t_{\alpha},
\end{equation}
so that the coefficients
\begin{eqnarray}
u_{\alpha}^{\prime} & = & \frac{1}{2}\left(s_{\alpha}+t_{\alpha}\right)/\sum_{\alpha}s_{\alpha}t_{\alpha}\label{eq:normu},\\
v_{\alpha}^{\prime} & = & \frac{1}{2}\left(s_{\alpha}-t_{\alpha}\right)/\sum_{\alpha}s_{\alpha}t_{\alpha}\label{eq:normv},
\end{eqnarray}
give spatial states $u^{\prime}(x)$ and $u^{\prime}(x)$ that obey
the normalization
\begin{equation}
\int dx\left(u^{\prime2}(x)-v^{\prime2}(x)\right)=1.
\end{equation}

The results of the numerical method used are shown in Figs. \ref{Rag1} to
\ref{Rag3}. In the expansion method, sufficient accuracy was often achieved
with a very small number (six) of basis states. This is not surprising
since we know that a two mode expansion in treating the GPE is quite
accurate as seen in Sec. \ref{subsec:Two-mode-approximation-for}. But we always checked the results with
larger number of basis states to be certain. The agreement with the
numerical method is excellent.

We noted in Fig. \ref{Rag3} that an oscillatory instability 
(complex $\lambda$)
arises for $\gamma > 5.94$ in the 
antisymmetric case for attractive interactions. Refs. \cite{Sac1,Sac2} do 
not report oscillatory instabilities, because this requires at least three states.
With just two states, the matrices $M_{1}$ and $M_{3}$ are already diagonal because of the 
symmetry of the wave functions makes $\Omega_{12}=\Omega_{21}=0$. 
$M_{1}M_{3}$ is then diagonal with real matrix elements that can be positive or negative. If one  
Is negative then $\lambda$ is imaginary, but never complex.   In the three state case,  
the matrix $M_{1}M_{3}$   has off-diagonal 
[1,3] and [3,1] elements. Now we have a 3x3 matrix with real elements that can have complex 
eigenvalues and $\lambda$ is then also complex.

\end{document}